\documentclass{ws-jai}

\usepackage[dvipsnames]{xcolor}

\begin{document}

\catchline{}{}{}{}{} 

\markboth{Young Min Seo}{Applications of Machine Learning Algorithms In Processing Terahertz Spectroscopic Data}

\title{Applications of Machine Learning Algorithms In Processing Terahertz Spectroscopic Data}

\author{Young Min Seo$^{1}$, Paul F. Goldsmith$^{1}$, Volker Tolls$^{2}$, Russell Shipman$^{3}$, Craig Kulesa$^{4}$, William Peters$^{4}$, Christopher Walker$^{4}$, Gary Melnick$^{2}$}

\address{
$^{1}$Jet Propulsion Laboratory, California Institute of Technology, 4800 Oak Grove Drive, Pasadena, CA, 91109, USA\\
$^{2}$Harvard–-Smithsonian Center for Astrophysics, 60 Garden Street, Cambridge, MA 02138, USA\\
$^{3}$SRON Netherlands Institute for Space Research, Landleven 12, 9747 AD Groningen, The Netherlands\\
$^{4}$Department of Astronomy \& Steward Observatory, University of Arizona, 933 N. Cherry Ave., Tucson, AZ 85721, USA
}

\maketitle

\corres{$^{1}$Young Min Seo, seo3919@gmail.com.}

\begin{history}
\received{(to be inserted by publisher)};
\revised{(to be inserted by publisher)};
\accepted{(to be inserted by publisher)};
\end{history}

\begin{abstract}

We present the data reduction software and the distribution of Level 1 and Level 2 products of the Stratospheric Terahertz Observatory 2 (STO2). STO2, a balloon-borne Terahertz telescope, surveyed star-forming regions and the Galactic plane and produced approximately 300,000 spectra. The data are largely similar to  spectra typically produced by single-dish radio telescopes.  However, a fraction of the data contained rapidly varying fringe/baseline features and drift noise, which could not be adequately corrected using conventional data reduction software. To process the entire science data of the STO2 mission, we have adopted a new method to find proper off-source spectra to reduce large amplitude fringes and new algorithms including Asymmetric Least Square (ALS), Independent Component Analysis (ICA), and Density-based spatial clustering of applications with noise (DBSCAN). The STO2 data reduction software efficiently reduced the amplitude of fringes from a few hundred to 10 K and resulted in baselines of amplitude down to a few K. The Level 1 products typically have noise of a few K in [CII] spectra and $\sim$1 K in [NII] spectra. Using a regridding algorithm, we made spectral maps of star-forming regions and the Galactic plane survey using an algorithm employing a Bessel-Gaussian kernel. The Level 1 and 2 products are available to the astronomical community  through the STO2 data server and the DataVerse. The software is also accessible to the public through Github. The detailed addresses are given in Section 4 of the paper on data distribution. 

\end{abstract}


\section{INTRODUCTION} \label{sec:intro}

Spectroscopic observations in the far-infrared (FIR) and submillimeter wavelengths have been critical to astronomy; for example, probing kinematics, tracing chemistry, and characterizing physical conditions in different phases of the interstellar medium. With the development of highly sensitive receivers and spectrometers at far-infrared and submillimeter wavelengths during the last decades, the amount of data has exploded and been bringing a wealth of new knowledge about our universe. On the other hand, processing, and publishing these extensive data sets has became a significant challenge. A robust, automated pipeline software that can handle a large amount of data and a wide range of spectral features is a critical element for successful future astronomical facility or mission.       

The Stratospheric Terahertz Observatory 2 (STO2) is one of the missions that has produced large data sets that challenge data processing, as a result of the volume and characteristics of the data. STO2 is a balloon-borne survey telescope, which observed the Galactic plane and several star-forming regions and produced over 300,000 spectra. While a \textcolor{black}{large fraction} of the data had no significant problems, a fraction of the data included rapidly-varying features and noise (e.g., fringes with the pattern changing significantly every few tens of seconds, with amplitude $>$50 K). \textcolor{black}{The cause of those features are still under investigation but most probable causes are electrical standing waves and LO power fluctuations. We found that the fast-varying features} could not be corrected efficiently using conventional data reduction algorithms \textcolor{black}{\citep{1981ApJ...250..341K,shipman17}}. Notably, algorithms that require prior information about source velocity to process the data are not suitable for the Galactic plane survey. Thus, processing STO2 data required adopting more flexible and robust algorithms\textcolor{black}{, including automatic identification of line emission during the reduction.}

To facilitate the data \textcolor{black}{processing}, considering \textcolor{black}{the} fraction of the data with low quality, we have employed several machine learning algorithms in the STO2 data processing pipeline. The machine learning algorithms have shown promising results in removing unwanted features in the STO2 data, including the high-amplitude fringes and mid-frequency drift noise. This paper elaborates on the algorithms in the STO2 data pipeline and their advantages and limitations.

In \S2 we briefly overview STO2 data, characterize STO2 data features, and discuss challenges in STO2 data reduction. In \S3, we elaborate on the algorithms for the STO2 data and the results. We briefly describe the Level 1 and 2 data products and their distribution in \S4. Finally, we discuss and summarize the limitations of the algorithms in \S5.

\section{STO2 Observation and Characteristics of STO2 Data} \label{STO_obs}

\begin{figure*}[tb]
\centering
\includegraphics[angle=0,scale=0.46]{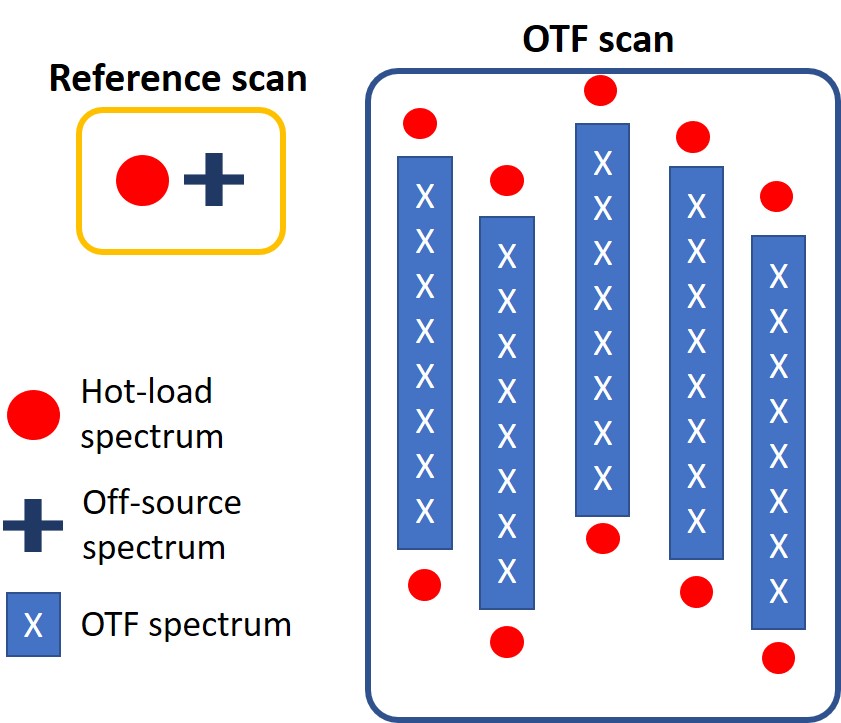}
\caption{On the fly (OTF) observation sequence of STO2. Every OTF observation of STO2 starts with a reference scan, which consists of taking hot (red disk) and off-source (cross) spectra, and then moving to the OTF scan. A single OTF scan leg consists taking on-source (OTF spectra, white crosses in blue rectangle) and hot (red disks) spectra. }
\label{obs_sequence}
\end{figure*}


The STO2 data are velocity resolved spectral observations of the [CII] \textcolor{black}{and}  [NII] lines recorded in 1024 channels with 1 MHz resolution, which is equivalent to a velocity resolution of 0.16 km s$^{-1}$ at 1.9 THz. The emission lines of [CII] and [NII] may have intrinsic shapes of a few to 20 km/s wide in complex line shapes in the STO2 survey. Each output spectrum is saved as a separate file in the Single-Dish-FITS format \citep{garwood00}, which is a binary table. We incorporate the scan number and the OTF dump number in their file names. We also include a summary of the observation information, for example, telescope telemetry data, observation position, integration time, and target names in the FITS headers.

\begin{figure*}[tb]
\centering
\includegraphics[angle=0,scale=0.7]{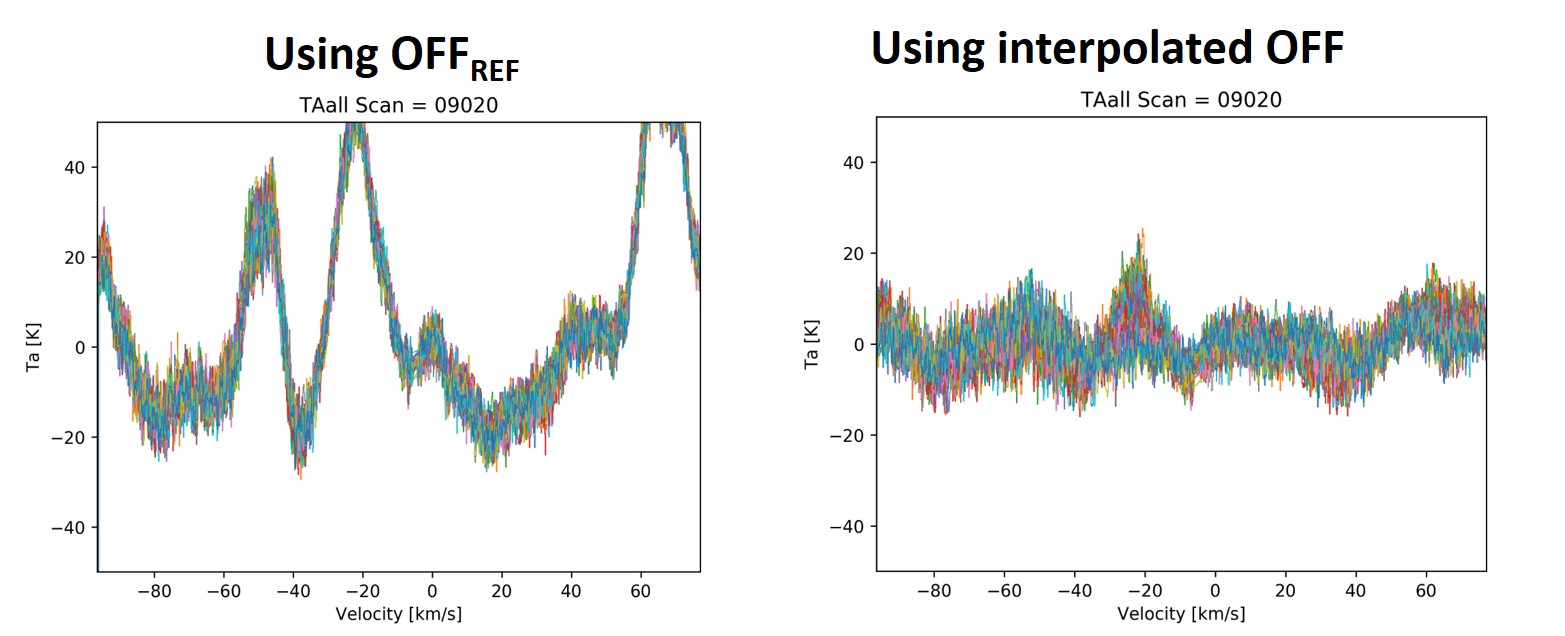}
\caption{Spectrum processed using the Off-source spectrum at a reference position (left) {\it vs.} spectrum processed using the interpolated Off-source spectra.}
\label{interp}
\end{figure*}

Scientific observations using STO2 were carried out in the On-The-Fly (OTF) mode. The typical OTF sequence of STO2 is described in Figure \ref{obs_sequence}. An OTF observation starts with a reference scan, which takes hot-load and off-source spectra at a reference position prior to starting the OTF scan and is used to estimate gain response at every channel. Hot-load spectrum are taken by observing an internal hot load on the gondola, which is at the ambient temperature within the payload. Reference positions are carefully selected not to have any significant [CII] and [NII] emission based on previous observations using ISO, Herschel, Spitzer, and ground-based radio telescopes. We also observed the hot load in the middle of the OTF scans if a single OTF scan lasted longer than 30 seconds to track noise and gain variations per channel over time. The \textcolor{black}{longest spectroscopic} Allan variance \textcolor{black}{minimum} time for STO2 is close to 30 seconds. \textcolor{black}{A single raster observation is typically in the range of 30 to 220 seconds, and consists of single or multiple sets of hot-load scans and a single OTF scan leg (typical single OTF scan leg duration < 30 seconds). In terms of number of files, it is equivalent to 45 -- 270 OTF file-outputs.} Typical integration times per OTF spectrum \textcolor{black}{and hot-load spectrum are 0.65 and 11 seconds, respectively.} Using a conventional data reduction method \citep[e.g.,][]{1981ApJ...250..341K}, we first assessed the characteristics of the STO2 data and found two challenging features in a fraction of STO2 data. \textcolor{black}{These} are the following:\\
1. The hot-load spectra in the reference scans are significantly different from the hot-load spectra taken during OTF scans (Figure \ref{interp}), \textcolor{black}{which might be due to temperature gradient or thermal instability of the telescope.} Having different hot-load spectra at different observation positions suggests that the gain of the receiver system varies with time and as a function of the observation pointing angle. It also indicates that we cannot directly use the spectra from the reference scans to calibrate the OTF observations. Due to this problem, significantly large fringes (amplitude of 20 -- 250 K) appear in the processed spectra if we use the hot-load and off-source spectra in the reference scans to calibrate OTF spectra.\\
2. Baselines of the spectra fluctuate over a short period ($<$30 seconds) \textcolor{black}{for a significant fraction of data}, which is likely due to electrical standing waves and LO power fluctuation. The baseline varies appreciably within one OTF scan leg. There can also be a sudden increase of the fringe amplitude followed by decay over time.\\

We could not achieve the desired science quality spectra using the conventional data-reduction method due to the above two features. Thus, we have developed an optimized data-processing algorithm for the STO2 data-reduction pipeline to obtain the best quality spectra. The flow of the STO2 data-reduction pipeline is shown in Figure \ref{fig_pipe}. The pipeline performs the conversion of the raw data into antenna temperature, de-fringing, baseline correction, and regridding. In the following subsections, we describe details of the STO2 data-reduction pipeline with elaborations of the methods to suppress the two features discussed above. 

\section{STO2 Data Reduction Software} \label{app:pipe}

\subsection{Conversion to the Antenna Temperature} \label{app:pipe:ta}
The first step of the STO2 data reduction is to convert the raw data in units of counts/s to antenna temperature. A conventional way for the conversion is to use the following equation for each spectral channel:
\begin{eqnarray}
T_{\rm A} = T_{\rm sys} \frac{{\rm OTF} - {\rm Off}_{\rm Ref}} {{\rm Off}_{\rm Ref}},
\label{ta_typical}
\end{eqnarray}
where $T_{\rm A}$, $T_{\rm sys}$, ${\rm OTF}$, and ${\rm Off}_{\rm Ref}$ denote antenna temperature, system temperature, spectrum from OTF scanning, and off-source spectrum at a reference position (reference scan), respectively. $T_{\rm sys}$ is estimated at the reference position. The equation can be applied as long as the gain of the spectrometer channels does not vary between the reference position and the on-source position. Unfortunately, STO2  \textcolor{black}{data exhibited} spectrum gains \textcolor{black}{that}vary significantly with the telescope pointing position between a reference observation and the OTF leg, and the baselines of off-source spectra are vastly different from those of the OTF spectra. The results using equation (\ref{ta_typical}) were not adequate for scientific use, as shown in the left panel of Figure \ref{interp}. 

To rectify the problem, we must find appropriate off-source spectra to correct the OTF scans. Using OTF spectra at the edge of a map where no emission is detected for an off-source spectrum is a frequently-used method when there are no reference spectra. However, for the STO2 mission, the target lines are [CII] and [NII], which typically extend across large areas and often stretch beyond the areas covered. Also, in a Galactic plane survey, the location and spatial extent of the emission are rarely known. Using spectra at the edges of maps is not suitable for STO2 data reduction. 

Therefore, we focus our effort on utilizing the hot-load spectra in reference scans and those obtained during OTF scans to predict off-source spectra for the OTF scans. With a detailed analysis of the hot-load and off-source spectra in reference and OTF scans, we found that the ratio of the hot-load spectrum to the off-source spectrum, Hot$_{\rm Ref}$/Off$_{\rm Ref}$, does not vary with the telescope pointing position and evolves relatively slowly over time. We also found that the hot-load spectra taken during OTF scans, Hot$_{\rm OTF}$, vary slowly over time and show the same gain as that of the OTF spectra. Using these two results, we obtained off-source spectra for the OTF scans and made a modified version of Equation (\ref{ta_typical}) for the STO2 data reduction as
\begin{eqnarray}
T_{\rm A} = T_{\rm sys}(t_{\rm OTF}){{\rm OTF} - {\rm Off}_{\rm Interp}(t_{\rm OTF}) \over {\rm Off}_{\rm Interp}(t_{\rm OTF})},
\label{ta_new}
\end{eqnarray}
where $t_{\rm OTF}$ denotes output time when an OTF spectrum is dumped and recorded to a storage. Off$_{\rm Interp}$($t_{\rm OTF}$) is a off-source spectrum for the OTF spectra at $t_{\rm OTF}$, which is evaluated as 
\begin{eqnarray}
{\rm Off}_{\rm Interp}(t_{\rm OTF}) = {{\rm Hot}_{\rm OTF}(t_{\rm OTF}) \over {\rm Hot}_{\rm Ref}(t_{\rm OTF})/{\rm Off}_{\rm Ref}(t_{\rm OTF})},
\label{eq_sky}
\end{eqnarray}
where Hot($t_{\rm OTF}$) denotes hot-load spectrum linearly interpolated to $t_{\rm OTF}$. The subscripts of Hot, OTF, and Ref denote the hot-load spectrum taken in OTF and at reference scans, respectively. Off$_{\rm Ref}$($t_{\rm OTF}$) is obtained by linearly interpolating the off-source spectra at a reference position to $t_{\rm OTF}$. T$_{sys}$($t_{\rm OTF}$) is the system temperature and defined as 
\begin{eqnarray}
T_{\rm sys}(t_{\rm OTF}) = {T_{\rm Hot} - y(t_{\rm OTF}) T_{\rm Off} \over y(t_{\rm OTF})-1},
\label{tsys}
\end{eqnarray}
where y($t_{\rm OTF}$) $\equiv$ Hot$_{\rm Ref}$($t_{\rm OTF}$)/Off$_{\rm Ref}$($t_{\rm OTF}$), $T_{\rm Hot}$ is the hot-load temperature (ambient temperature of the payload), which is measured during reference scan and recorded in the FITS header, and \textcolor{black}{$T_{\rm Off}$} is the noise temperature of an off-source (empty) sky, which is roughly 45 K at the altitude of STO2. This method resulted in a significant improvement as shown in the right panel in Figure \ref{interp}. \textcolor{black}{The typical Single Sideband T$_{sys}$ ranges from 3000 K to 3800 K at 1.9 THz.}

There are some OTF observations that do not have any hot-load observations, which makes it inappropriate to use Equation (\ref{eq_sky}). For the observations that do not have a hot-load spectrum within an OTF scan, we use a library of the ratios of hot-load spectra, which is a collection of all observed Hot$_{\rm OTF}$/Hot$_{\rm Ref}$ at every observation time-step of Hot$_{\rm OTF}$. 
We interpolated the Hot$_{\rm Ref}$ to obtain the Hot$_{\rm Ref}$ at the time-step of Hot$_{\rm OTF}$. Using the library, we create a series of the predicted off-source spectra at time $t$ as
\begin{eqnarray}
{\rm Off}_{\rm lib}(t;i) \equiv {{\rm Off}_{\rm Ref}(t)\over ({\rm Hot}_{\rm OTF}/{\rm Hot}_{\rm Ref})_i},
\label{library}
\end{eqnarray}
where the subscript $i$ denotes $i$-th element of the library, Off$_{\rm lib}(t;i)$ is the off-source spectrum evaluated at time $t$ using the $i$-th element of the library. With the series of predicted Sky spectra, we reduce the first and the last OTF spectra within a single OTF scan and choose the best off-source spectrum that results in the flattest baseline. We found that the method of using the library gives better results than using the conventional method \citep[e.g.,][]{1981ApJ...250..341K}, but not as good as those obtained using the interpolation method using Equation (\ref{ta_new}). \textcolor{black}{This method is similar to the data reduction method of the \textit{Herschel} HIFI instrument, which extracts the families of baseline from Off-position spectra based on the spectral features and selects the optimal baseline for spectra using a Bayesian approach. The same analysis of the STO2 data reduction revealed that there are more than 1,000 families of baselines in the STO2 spectra, while the spectra of the HIFI instrument have much smaller number of families ($<$ 100, HIFI Data Reduction Guide\footnote{https://www.cosmos.esa.int/web/herschel/legacy-documentation-hifi}). We found that the method was not efficient for STO2 data due to the high diversity of features in the STO2 data. }

\subsection{Spectrum Quality Assessment \& De-fringing} \label{app:pipe:base}

\begin{figure*}[tb]
\centering
\includegraphics[angle=0,scale=0.46]{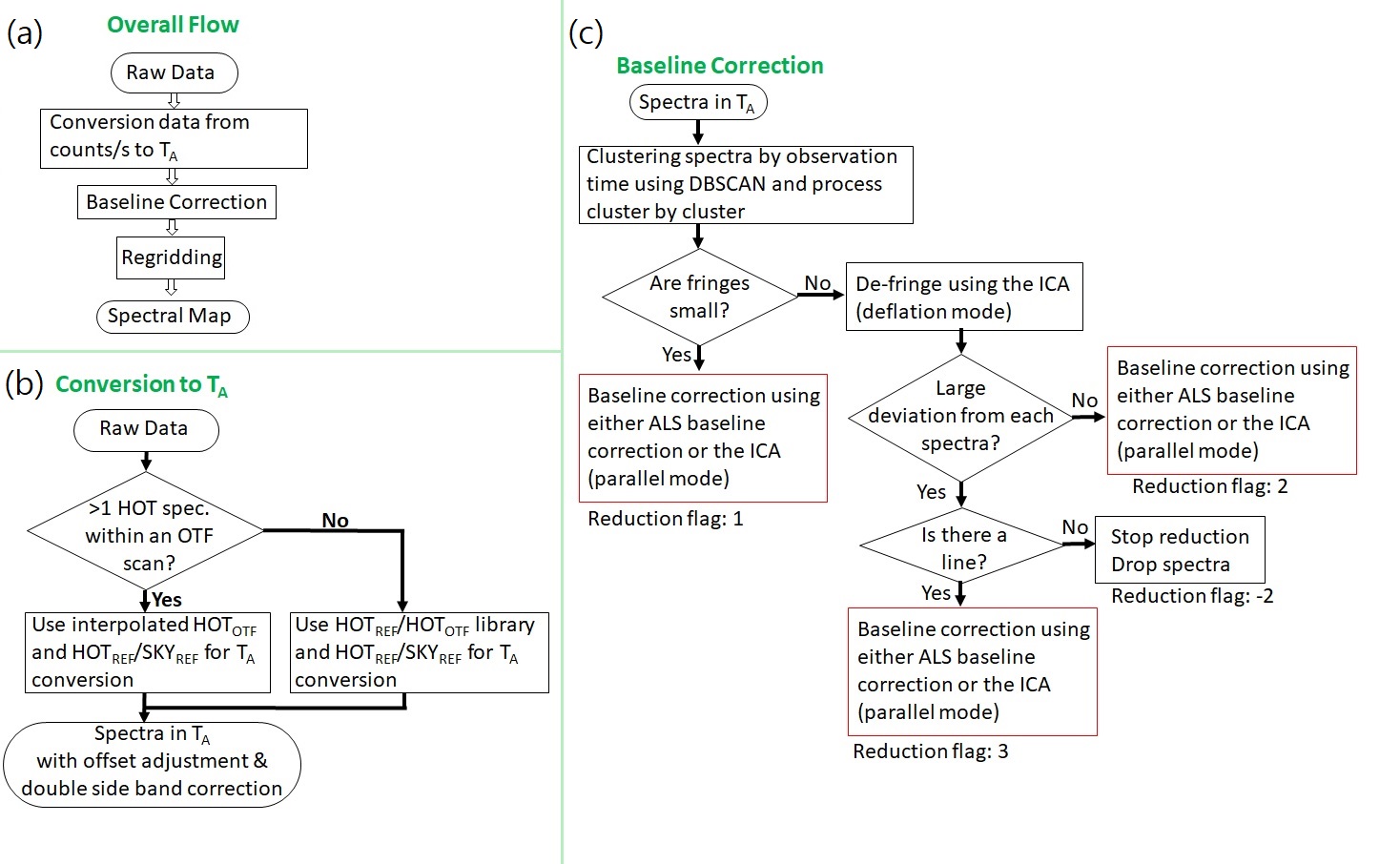}
\caption{Flow charts of the STO2 data reduction pipeline. Panel (a) shows the overall software flow. Panel (b) elaborates on converting the raw data to the data in units of antenna temperature. Panel (c) describes the flow of the fringe and baseline correction.}
\label{fig_pipe}
\end{figure*}

\begin{figure*}[tb]
\centering
\includegraphics[angle=0,scale=0.69]{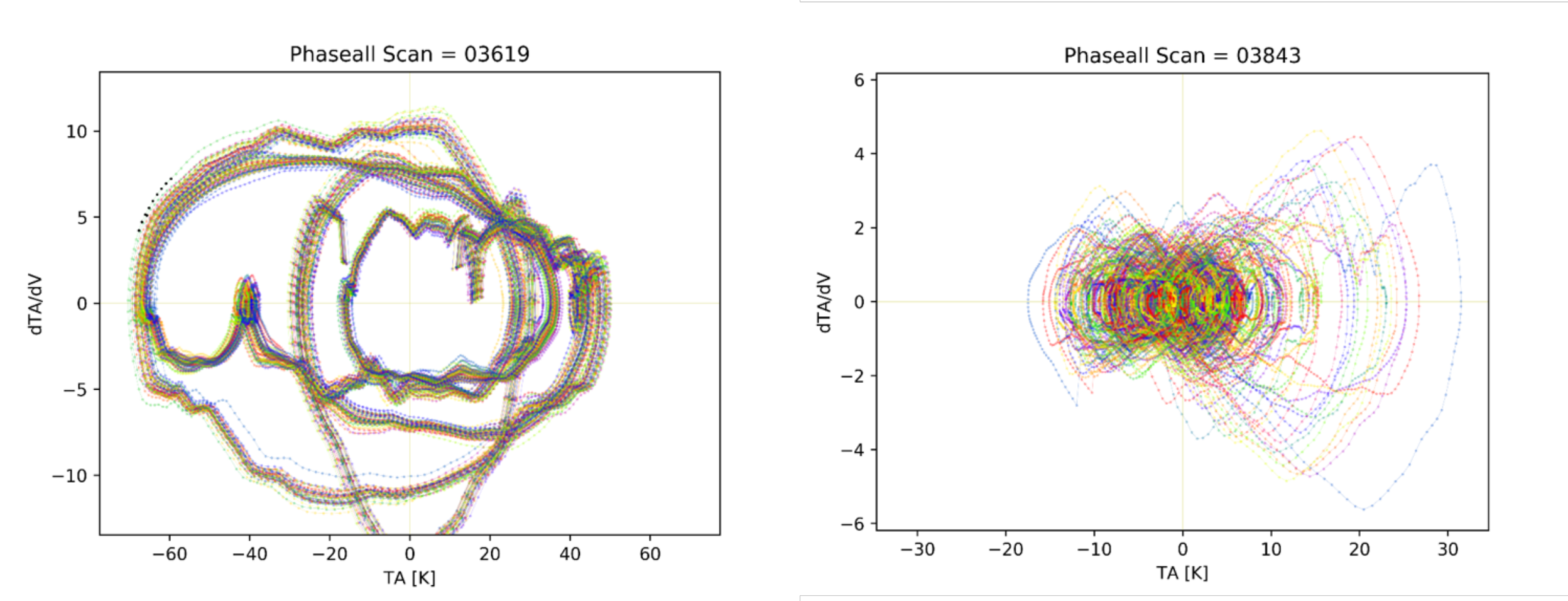}
\caption{Phase portrait of de-noised spectra in Scan 3619 (left) and Scan 3843 (right). Each dot corresponds to values of one channel. The spectra in Scan 3619 have a large fringe amplitude while the spectra in Scan 3843 have small fringes as well as clear [CII] lines, which is shown as large fan-shaped features in the positive Ta quadrants.}
\label{fig_pp}
\end{figure*}

In the second step of the STO2 data reduction, we group spectra based on the observation time, assess spectrum quality and remove fringes. We found that fringe patterns are similar to each other when the spectra are taken within a single OTF scan (typically $<$30 seconds or $<$40 outputs). Based on this characteristic, we first cluster the spectra using observation time as a main parameter of the clustering procedure. Among many available clustering algorithms, we adopt the Density-Based Spatial Clustering of Applications with Noise (DBSCAN) in the Scikit-Learn \citep{scikit-learn}, which is a machine learning package in python. We use DBSCAN because DBSCAN automatically clusters a large number of OTF spectra ($>$300,000) with inhomogeneous OTF duration and variations in the observation intervals due to failed observations, telescope slews, and reference observations. We adjust the parameters of DBSCAN to cluster all OTF spectra neighboring within a $<$2-second interval as one group.  

After DBSCAN designates the group number for spectra, we process spectra in a group by the decision tree shown in Figure \ref{fig_pipe}. We process spectra \textcolor{black}{group by group} because the STO2 spectra have quite different fringe patterns after each reference observation or \textcolor{black}{hot-load} spectrum. In the decision tree, we evaluate the data quality and fringe patterns (e.g. fringe amplitudes and line strength) and determine optimized processes for de-fringing and baseline correction.

We use three parameters to evaluate the quality of spectra: average of the squared deviation, standard deviation of the square deviation, the maximum density position in the phase portrait. The squared deviation is defined as 
\begin{eqnarray}
P_i \equiv \sum\limits_{v_k=v_1}^{v_k=v_2} T_i^2(v_k), 
\end{eqnarray}
where $P_i$ is the squared deviation of the $i$-th spectrum in a group of the OTF spectra, $v$ is velocity, and $T_i$ is the antenna temperature of the $i$-th spectrum. The subscripts 1 \& 2 are the minimum and the maximum velocity range of the spectrum, which is set to cover the half of spectral range. We make the phase portrait of spectra, which is a plot of $T(v)$ {\it vs.} $\partial T/\partial v$, after we substantially reduce the white noise using a wavelet de-noising method within each spectrum. For a spectrum with large fringes, the square deviation is significantly larger than the squared deviation of a spectrum with no fringes. Also, the maximum density point in the phase portrait deviates considerably from (0,0) (Figure \ref{fig_pp}). The standard deviation of the squared deviation indicates whether the fringes of spectra within a group are similar or vary considerably within a group, for example, a group with similar baselines found to have values less than 10$^5$, while a group with fastly varying baselines found to have typically over 10$^6$. For a group of spectra with strong lines but without larges fringes, the value of squared deviations is typically small and the maximum density point is close to (0,0) in the phase portrait but there is a large standard deviation in the squared deviations of the spectra. Thus, using these values, we classify the groups into the four different categories (Figure \ref{fig_pipe}) and determine which algorithm to process the group of spectra.

The next step is the de-fringing for a group of spectra with significantly large fringes (fringes with  amplitudes $>$30 K, reduction flags 2, 3, and -2). We skip this process for a group without large fringes (reduction flag 1). We use the independent component analysis (ICA) to do the de-fringing. The ICA algorithm is often used to decompose different sources within a signal assuming that the observed signal is linearly mixed from multiple sources. We assume that fringes are the combinations of different contributions to frequency-dependant output (e.g., optical and electrical standing waves) in the STO2 system. Among many modes of ICA, we use the deflation mode of ICA (deflation ICA). The deflation ICA delivers ICA-components in an orderly manner based on amplitude, and the first ICA-component of the deflation ICA is a feature with the largest amplitude. Thus, we use only the first ICA-component to correct the fringe of the largest amplitude throughout a group. We found that residual fringes are typically $\pm$10 K after correction. This method is efficient when the amplitude of fringes is larger than that of line emission in all spectra within a single group.  

We found that the first ICA-component can be contaminated by the line emission in the spectra when the amplitude of line emission is comparable to that of the strongest fringe. To check whether or not the first ICA-component is contaminated by strong emission, we examine the $rms$ of each channel across spectra within a group after de-fringing. We found that the channels involving strong emission typically show significantly larger $rms$ values compared to those of the other channels when the emission varies within a group (reduction flag 3). For such a group, we mask the channels \textcolor{black}{that have large $rms$ values} and re-do the de-fringing step. The masked channels are replaced by a second-order polynomial curve. We found that this additional procedure prevents removing the line emission during the de-fringing step, but this method only works when the intensity or shape of the emission varies significantly within a group. If the line shape and intensity are the same throughout the spectra within a group, this method will fail to isolate the line emission from fringes. We also found that there are groups with all channels having large $rms$ after de-fringing (reduction flag of -2). In such groups, fringes vary rapidly, typically in less than 1 second, and the deflation ICA failed to find a common fringe pattern for the group. We stopped further processing of these groups since we could not process them using either conventional methods (e.g., high-order polynomial fitting) or the deflation ICA.

\subsection{Baseline Correction using the ALS and the Parallel ICA and Regridding} \label{app:pipe:pICA}

In the third step of the STO2 data reduction, we correct the baselines of the de-fringed spectra and the spectra with small fringes. We correct the baselines using either the asymmetric least square method (ALS, \citealt{eilers05}) or the ICA in the parallel mode (parallel ICA) or both. The ALS baseline correction efficiently removes any low-frequency fringe. The parallel ICA can remove both low- and high-frequency fringes simultaneously but tends to be slower compared to the ALS baseline correction. The parallel ICA is different from deflation ICA in that it evaluates features without ordering the amplitude of features. In signal processing, the parallel ICA is often used for the blind source separation when an output signal is linearly composed of many input sources \citep{jain12}. We assume that the spectra of STO2 are a mixture of standing waves from the receiver electronics and the emission from astronomical sources, we separate the fringes and the emission using the parallel ICA. In the pipeline, we use the ALS followed by the parallel ICA as a default setting for the baseline correction. After baseline correction, we found that the residual fringes are typically $\sim$1 K. The default baseline correction using the ALS and the parallel ICA has the following four steps:

\begin{figure*}[tb]
\centering
\includegraphics[angle=0,scale=0.1]{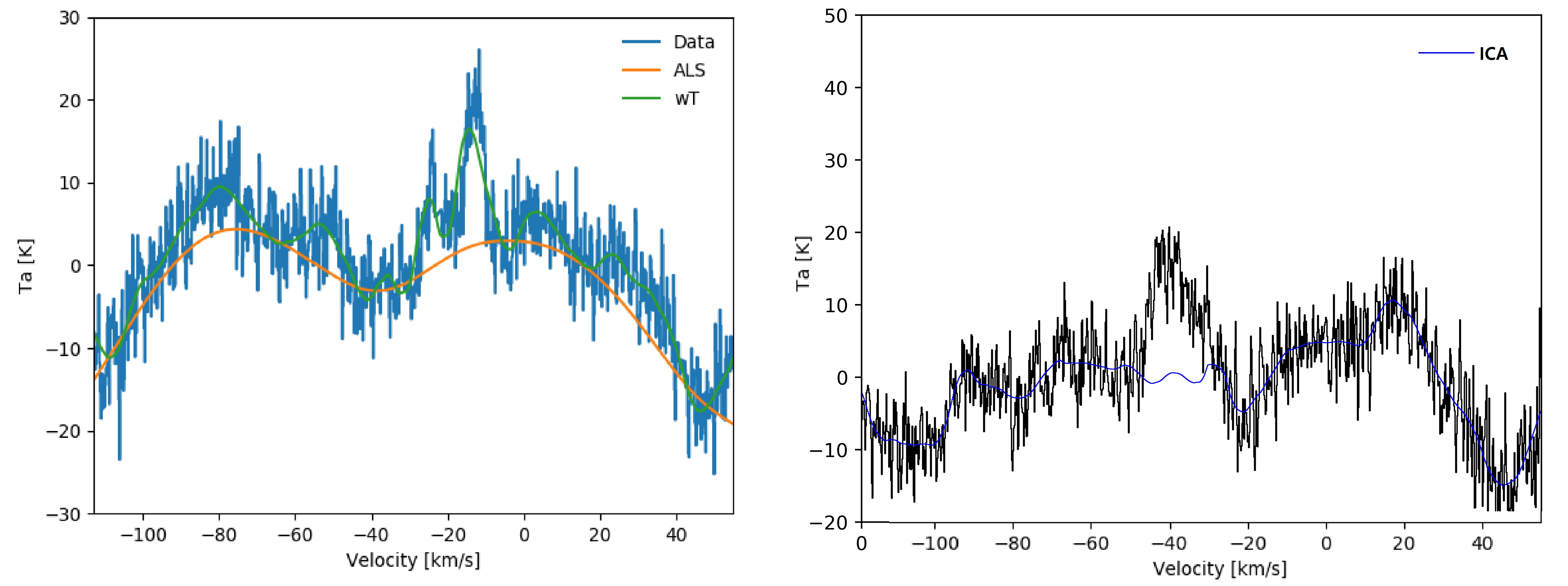}
\caption{Baseline correction using a wavelet de-noising method and the asymmetric least square (left) and the independent component analysis (right). }
\label{base}
\end{figure*}

1. We remove the white noise in the spectra using a wavelet de-noising algorithm. We also eliminate low-frequency fringes/fluctuations using the ALS baseline correction to all de-noised-spectra within a group. This step is selected in the default setting but can be turned off.

2. We apply the parallel ICA to denoised-spectra to further correct high/mid-frequency fringes. We generate $N$ ICA-components as a default setting if a group of spectra has $N$ spectra. It is possible to generate an arbitrary number of ICA-components, but if the number of the ICA-components is too small ($<$5), we found that the astronomical line emission and fringes are not adequately separated. If we generate more components than $N$ components, the parallel ICA provides a better separation of fringes from the line emission in exchange of computational costs.          

3. The third step is to remove any contribution originating from the line emission in all ICA-components. We found that the parallel ICA decomposes the fringes into $N$ components, but it also decomposes line emission into a few ICA-components rather than concentrating it in a single ICA-component. We determine the contribution of line emission within ICA-components using a combination of the sigma-clipping, the peak detection algorithm, and the Gaussian-line-fitting algorithm. After line detection, we replace the contribution of the line emission in each ICA-component with a low-degree polynomial (the default is a linear function).

4. We construct the baselines by combining the modified ICA-components using the original mixing matrix obtained by the parallel ICA and correct the baseline of the spectra.

Fringes with either a broader or narrower width than the FWHM of line emission are efficiently removed using this method (Figure \ref{base}). But if fringe width is similar to the FWHM of the line emission and the amplitude of fringes are higher than the line amplitude, this method may not isolate the baseline and fringes from the line emission. 

To make maps from the STO2 spectra, we need to re-grid spectra. We use a regridding algorithm from \citet{mangum07}. We used a Gaussian-Bessel kernel for the regridding, as suggested in that paper.

\begin{figure*}[tb]
\centering
\includegraphics[angle=0,scale=0.65]{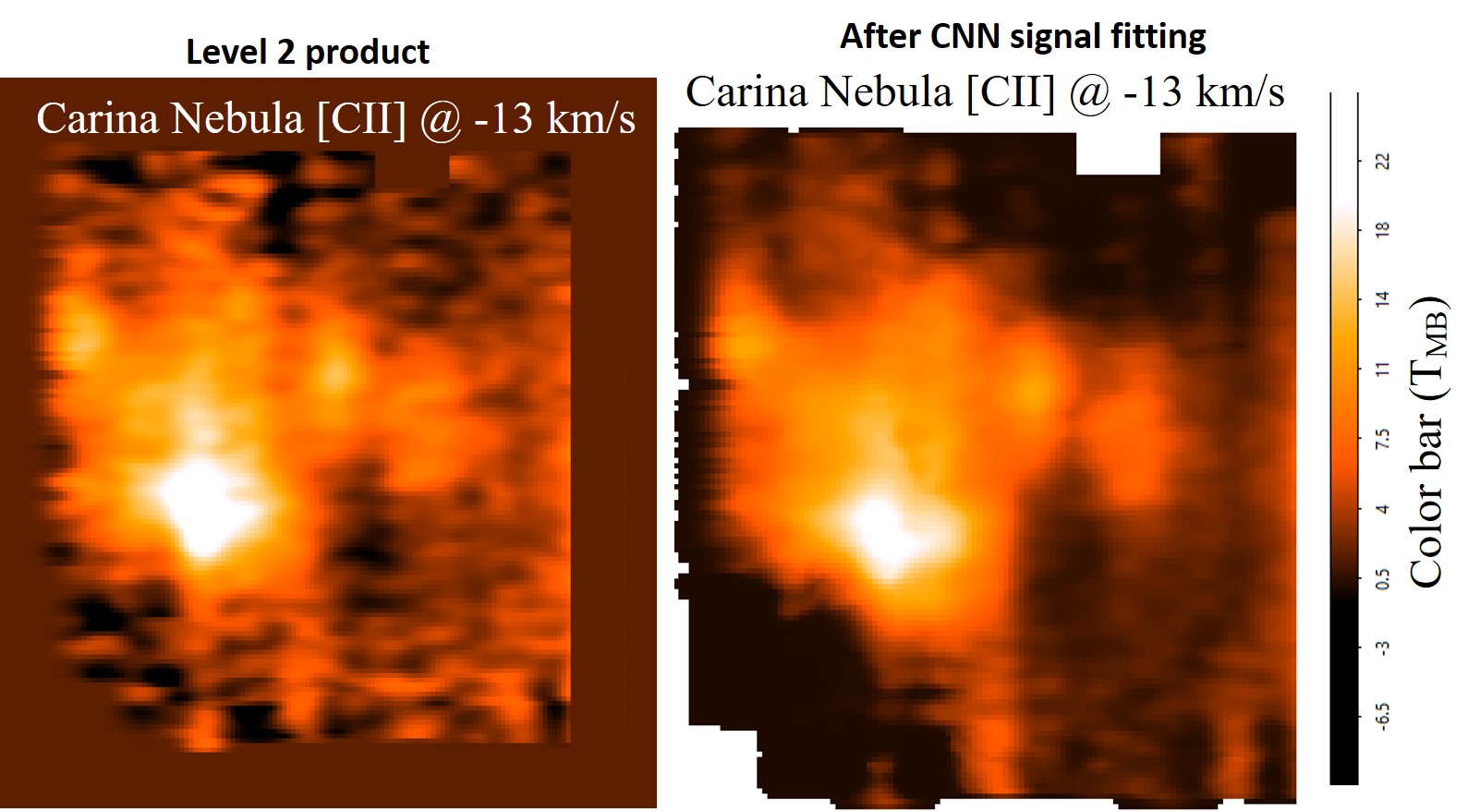}
\caption{Original [CII] map of Trumpler 14 region observed using STO2 data processing software (left, \citet{seo19}) and signal-fitted map using a convolutional neural network (right). }
\label{CNN}
\end{figure*}

\subsection{Signal Fitting Using a Convolutional Neural Network} \label{app:pipe:CNN}

The last step in the STO2 data reduction is a signal-fitting software using a convolutional neural network (CNN), more specifically, CNN autoencoder. The STO2 data contains surveys of multiple high-mass star-forming regions, which have complex velocity structures. Thus, the line profiles in STO2 data are very complex and require multiple Gaussian functions to fit a line profile. A conventional method to fit a line profile employs least $\chi^2$ fitting and is useful if a line profile can be fitted using fewer than five Gaussian functions. If more than five are required, the number of fitting parameters becomes too large, and the least $\chi^2$ method takes an excessively long time to obtain a proper solution. On the other hand, CNN is excellent in recognizing and processing highly complicated shapes or patterns. Line profiles in astronomy are often a series of superposed Gaussian functions. We generate a training set containing noise, fringes, and signals, and we trained the CNN to eliminate noise and fringes and to return only the signal. The training set includes 500,000 samples with various combinations of simulated baseline (sine waves and polynomials) and signals (up to 20 skewed Gaussian functions). \textcolor{black}{In testing} the CNN signal-fitting through a simple Monte-Carlo simulation, we found that roughly 90\% of 2$\sigma$ emission is detected and fitted correctly, whereas the CNN is not able to recover any emission weaker than 1.5$\sigma$. We applied the CNN to \textcolor{black}{ the spectral maps of the STO2 observations (Figure \ref{CNN})} and found that fitting signals is significantly faster than a conventional Gaussian fitting algorithm, and we could process 10,000 spectra, \textcolor{black}{including complex line profiles with more than several intensity peaks,} in less than one second with highly accurate results.    

We have also written a CNN for baseline correction, which is not included in the STO2 pipeline. We found that it delivers results similar to or better than the ICA de-fringing and baseline correction in many cases and is at least 100 times faster in computation speed. However, we found that it is difficult to generate a training set covering all features in the STO2 data since the STO2 features are \textcolor{black}{extremely} diverse, having more than 1,000 families of features in ~300,000 spectra. The algorithm will be further tested using SOFIA \textcolor{black}{GREAT} data.   

\begin{figure*}[tb]
\centering
\includegraphics[angle=0,scale=0.62]{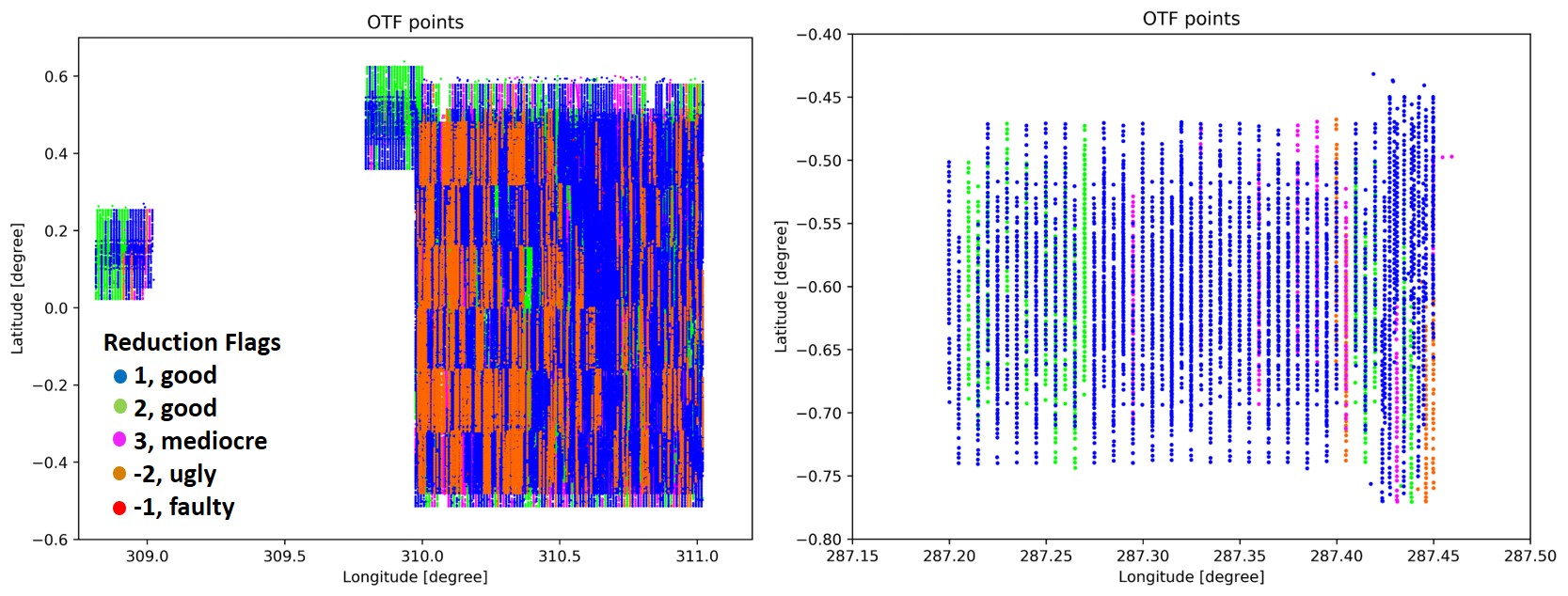}
\caption{Reduction flags of the STO2 Level 1 data. Each point denotes a single spectrum. Each flag is marked by different color with short descriptors (see text for details). }
\label{level1}
\end{figure*}

\section{Data Products and Distribution} \label{distribution}

We reduced all [CII] spectra of the STO2 science observations using the software described above. STO2 also observed [NII], but we found that the [NII] data do not reveal any  \textcolor{black}{appreciable ($>$ 5 K) [NII] emission towards sources expected to be at least this strong based on spectrally unresolved data and estimates of the line width},  due to an instrumental malfunction which is still being investigated. The [CII] Level 1 products are all the individual spectra processed through the software. Figure \ref{level1} shows the distribution of the reduction flags in the Level 1 product. We found that roughly 40\% of the spectra show relatively weak fringes and are designated as reduction flag 1. These are relatively good spectra, which required only baseline corrections. The final $rms$ values for these spectra are typically $\leq$5 K at a velocity resolution of 1 km s$^{-1}$\textcolor{black}{, while the expected $rms$ noise is roughly 2 K if the instrument follows the radiometer equation.} Another 40\% of the spectra have large-amplitude fringes, which were corrected through the de-fringing step using the deflation ICA method. These are flagged as 2 and 3. The spectra so flagged as 2 and 3 showed $rms$ values $\sim$6 K, lower quality than those flagged as 1. The Level 2 products are spectral maps by re-gridding the Level 1 products. The Level 2 products include three star-forming regions, Trumpler 14, Gum 31, and NGC 3576, and three Galactic plane regions at $l$ = 286$^\circ$, 310$^\circ$, and 328$^\circ$, with $|b|$ $\leq$ 0.5$^\circ$. The STO2 observations of the Trumpler 14 region have been analyzed in detail and published in \citet{seo19}.

The validation of the final reduction is carried out for Trumpler 14 map using an integrated intensity map of the same region from Infrared Space Observatory (ISO). We found that only ISO observed the same region, providing data that can be used for comparison to the STO2 observations. We found that the two maps have a good agreement (uncertainty of $\pm38\%$ in the integrated intensity) assuming the beam efficiency of STO2 to be 0.7 (Appendix 1 in \citealt{seo19}). The Herschel Space Observatory also observed this region, but the observations were carried out in a strip observation mode near the Carina I cloud in Tr 14, where there is a PDR region with an edge-on \textcolor{black}{orientation}. This PDR feature is a \textcolor{black}{thin transition layer at the molecular cloud surface} that is expected to span only several arcseconds, so spectra toward this region depend \textcolor{black}{sensitively} on the beam size of the telescope. We found that the spectrum from Herschel shows 80 K peak intensity in the line while STO2 shows 18 K peak intensity at the same position, which is reasonable since the beam size of STO2 is roughly four times larger than that of Herschel.          

Level 1 and 2 data products of STO2 surveys are available to the public and are accessible from the PI team server \footnote{\url{http://soral.as.arizona.edu/STO2/}} and DataVerse. The entire STO2 data processing software is written in python 3 and the scripts are publicly accessible from Youngmin Seo's Github\footnote{\url{https://github.com/seoym3919/STO2_PIPELINE}}.   

\section{Summary} \label{summary}
The STO2 mission, the first \textcolor{black}{successful} terahertz heterodyne balloon-borne \textcolor{black}{astronomical} observatory, was launched from the McMurdo station in December 2016. The mission was successful in surveying several star-forming regions and the Galactic plane, but the data required an enhanced effort to process due to unexpected features in spectra, including fringes with various frequencies and amplitudes, the dependence of fringe patterns on telescope position, and fast-varying fringe patterns. We found that it is inefficient to process $\sim$300,000 spectra using a conventional method. We have thus written advanced algorithms to suppress the characteristic fringes in the STO2 data and to produce scientifically usable data. The key efforts and results of the STO2 data-process software are as follow:

1. We found that the spectral output depends on the telescope position. The spectra at reference positions have vastly different form than the OTF spectra, and could not be used for converting the raw data to units of antenna temperature. To rectify the problem, we interpolated off-source spectra at reference positions using the ambient spectra at reference and on-source positions. The method efficiently reduced the amplitude of fringes from a few hundred to 10 K.

2. We found that the feature patterns vary rapidly, but are similar within a single OTF raster scan. We used machine learning algorithms to cluster spectra in each OTF scan, assessed spectrum quality and features, and applied optimized algorithms, including Independent Component Analysis (ICA) and Asymmetric Least Squares (ALS), to process fringes and standing waves. We found that roughly 40\% of the entire [CII] spectra are good (flagged as 1), which only needed baseline correction using either ICA or ALS. Another 40\% out of the spectra have large fringes with amplitudes over 50 K, which requires an additional step of de-fringing before baseline correction (flagged as 2 and 3). The quality of the processed spectra is typically worse than that of the spectra flagged as 1. For the last 20\% of the spectra, we found that the STO2 software could not efficiently process the spectra to a scientifically usable level. Thus, only the spectra flagged 1, 2, and 3 have been used for creating maps.

3. We found that the signal-fitting algorithm using a Convolutional Neural Network (CNN) autoencoder is extremely fast in fitting line profiles compared to the method minimizing $\chi^2$. We have also written a CNN autoencoder for correcting the baseline. We found that it is very effective and rapid to process a portion of the STO2 data but was not capable to process the entire data set since the fringe patterns vary excessively in parts of the STO2 data, which makes it hard to build an effective set of training models for the neural network. The algorithm may still be very useful for other data and will be tested further.      

4. Level 1 and 2 data products of the STO2 surveys are open to the public and are accessible from the PI team server (link in \S\ref{distribution}) and DataVerse. The entire STO2 data process software is written in python 3 and the scripts are publicly accessible from Youngmin Seo's Github (link in footnote c of \S\ref{distribution}). 

The STO2 mission serves as a pathfinder for future suborbital missions using balloon platforms such as GUSTO (NASA Explorer MO, PI: Chris Walker) and ASTHROS (NASA APRA, PI: Jorge Pineda). The software written for STO2 will be available to the two missions if an advanced data processing pipeline is required. 

\section{Acknowledgement}
We thank the anonymous referee for helping improve the paper in a variety of ways. We acknowledge that this work is supported by NASA Astrophysics and Data Analysis Program (17-ADAP17-0048). STO2 is a multi-institutional effort funded by the National Aeronautics and Space Administration (NASA) through the ROSES-2012 program under grant NNX14AD58G. This work was carried out in part at the Jet Propulsion Laboratory, which is operated for NASA by the California Institute of Technology.
\bibliographystyle{ws-jai}
\bibliography{STO2_pipe}

\begin{thebibliography}{8}
\newcommand{\enquote}[1]{``#1''}
\providecommand{\natexlab}[1]{#1}
\providecommand{\url}[1]{\texttt{#1}}
\providecommand{\urlprefix}{URL }
\expandafter\ifx\csname urlstyle\endcsname\relax
  \providecommand{\doi}[1]{doi:\discretionary{}{}{}#1}\else
  \providecommand{\doi}{doi:\discretionary{}{}{}\begingroup
  \urlstyle{rm}\Url}\fi

\bibitem[{Eilers \& Boelens(2005)}]{eilers05}
Eilers, P. \& Boelens, H. [2005]  \emph{Unpubl. Manuscr} .

\bibitem[{{Garwood}(2000)}]{garwood00}
{Garwood}, R.~W. [2000]  \enquote{{SDFITS: A Standard for Storage and
  Interchange of Single Dish Data},}  \emph{Astronomical Data Analysis Software
  and Systems IX}, eds. {Manset}, N., {Veillet}, C. \& {Crabtree}, D., p. 243.

\bibitem[{Jain \& Rai(2012)}]{jain12}
Jain, S. \& Rai, D. [2012]  \emph{IJEST} \textbf{4}.

\bibitem[{{Kutner} \& {Ulich}(1981)}]{1981ApJ...250..341K}
{Kutner}, M.~L. \& {Ulich}, B.~L. [1981]  \emph{Astrophysical Journal}
  \textbf{250},  341, \doi{10.1086/159380}.

\bibitem[{{Mangum} \emph{et~al.}(2007){Mangum}, {Emerson} \&
  {Greisen}}]{mangum07}
{Mangum}, J.~G., {Emerson}, D.~T. \& {Greisen}, E.~W. [2007]  \emph{Astronomy
  \& Astrophysics} \textbf{474},  679, \doi{10.1051/0004-6361:20077811}.

\bibitem[{Pedregosa \emph{et~al.}(2011)Pedregosa, Varoquaux, Gramfort, Michel,
  Thirion, Grisel, Blondel, Prettenhofer, Weiss, Dubourg, Vanderplas, Passos,
  Cournapeau, Brucher, Perrot \& Duchesnay}]{scikit-learn}
Pedregosa, F., Varoquaux, G., Gramfort, A., Michel, V., Thirion, B., Grisel,
  O., Blondel, M., Prettenhofer, P., Weiss, R., Dubourg, V., Vanderplas, J.,
  Passos, A., Cournapeau, D., Brucher, M., Perrot, M. \& Duchesnay, E. [2011]
  \emph{Journal of Machine Learning Research} \textbf{12},  2825.

\bibitem[{{Seo} \emph{et~al.}(2019){Seo}, {Goldsmith}, {Walker}, {Hollenbach},
  {Wolfire}, {Kulesa}, {Tolls}, {Bernasconi}, {Kavak}, {van der Tak},
  {Shipman}, {Gao}, {Tielens}, {Burton}, {Yorke}, {Young}, {Peters}, {Young},
  {Groppi}, {Davis}, {Pineda}, {Langer}, {Kawamura}, {Stark}, {Melnick},
  {Rebolledo}, {Wong}, {Horiuchi} \& {Kuiper}}]{seo19}
{Seo}, Y.~M., {Goldsmith}, P.~F., {Walker}, C.~K., {Hollenbach}, D.~J.,
  {Wolfire}, M.~G., {Kulesa}, C.~A., {Tolls}, V., {Bernasconi}, P.~N., {Kavak},
  {\"U}., {van der Tak}, F. F.~S., {Shipman}, R., {Gao}, J.~R., {Tielens}, A.,
  {Burton}, M.~G., {Yorke}, H., {Young}, E., {Peters}, W.~L., {Young}, A.,
  {Groppi}, C., {Davis}, K., {Pineda}, J.~L., {Langer}, W.~D., {Kawamura},
  J.~H., {Stark}, A., {Melnick}, G., {Rebolledo}, D., {Wong}, G.~F.,
  {Horiuchi}, S. \& {Kuiper}, T.~B. [2019]  \emph{Astophysics Journal}
  \textbf{878}, 120, \doi{10.3847/1538-4357/ab2043}.

\bibitem[{{Shipman} \emph{et~al.}(2017){Shipman}, {Beaulieu}, {Teyssier},
  {Morris}, {Rengel}, {McCoey}, {Edwards}, {Kester}, {Lorenzani}, {Coeur-Joly},
  {Melchior}, {Xie}, {Sanchez}, {Zaal}, {Avruch}, {Borys}, {Braine}, {Comito},
  {Delforge}, {Herpin}, {Hoac}, {Kwon}, {Lord}, {Marston}, {Mueller}, {Olberg},
  {Ossenkopf}, {Puga} \& {Akyilmaz-Yabaci}}]{shipman17}
{Shipman}, R.~F., {Beaulieu}, S.~F., {Teyssier}, D., {Morris}, P., {Rengel},
  M., {McCoey}, C., {Edwards}, K., {Kester}, D., {Lorenzani}, A., {Coeur-Joly},
  O., {Melchior}, M., {Xie}, J., {Sanchez}, E., {Zaal}, P., {Avruch}, I.,
  {Borys}, C., {Braine}, J., {Comito}, C., {Delforge}, B., {Herpin}, F.,
  {Hoac}, A., {Kwon}, W., {Lord}, S.~D., {Marston}, A., {Mueller}, M.,
  {Olberg}, M., {Ossenkopf}, V., {Puga}, E. \& {Akyilmaz-Yabaci}, M. [2017]
  \emph{Astronomy \& Astrophysics} \textbf{608}, A49,
  \doi{10.1051/0004-6361/201731385}.

\end{thebibliography}

\end{document}